\newcommand{\dis}{\displaystyle} 
\newcommand{\mc}{\cal}
\documentstyle[12pt]{article}
\begin{document}
\parindent 1.3cm
\thispagestyle{empty}   
\vspace*{-3cm}
\noindent

\def\arccot{\mathop{\rm arccot}\nolimits}
\def\sd{\strut\displaystyle}

\begin{obeylines}
\begin{flushright}
UG-FT-79/97
hep-ph/9707299
July 1997
\end{flushright}
\end{obeylines}
\vspace{2cm}

\begin{center}
\begin{bf}
\noindent
CONTACT TERMS IN CHARGED CURRENT PROCESSES AT HERA
\end{bf}
  \vspace{1.5cm}\\
FERNANDO CORNET and JAVIER RICO \footnote{E-mail addresses: cornet@ugr.es;
jrico@rigoberto.ugr.es}
\vspace{0.1cm}\\
Departamento de F\'\i sica Te\'orica y del Cosmos,\\
Universidad de Granada, 18071 Granada, Spain\\
   \vspace{2.2cm}

{\bf ABSTRACT}

\parbox[t]{12cm}{
We obtain bounds on the mass scales characterizing four-fermion contact 
interactions in charged current processes. The bounds arise
from the $Q^2$ and $x$ distributions in the processes 
$e^{\pm}p \to \nu^{^{\hbox{\hspace*{-3mm}{\tiny (---)}}}} X$ measured
by the two HERA experiments, H1 and ZEUS.}
\end{center}
\vspace*{3cm}

\newpage
The two HERA experiments (H1 and ZEUS) have published determinations of 
the $W$ mass from the measurement of the $Q^2$ distributions in charged 
current deep inelastic scattering processes 
$e^{\pm}p \to \nu^{^{\hbox{\hspace*{-3mm}{\tiny (---)}}}} X$. The results
they have obtained:
\begin{displaymath}
   \begin{array}{ll}
M_W = 84^{+10}_{-7} \; GeV & \hbox{H1 \cite{H1}} \\
M_W = 79^{+8+4}_{-7-4} \; GeV & \hbox{ZEUS \cite{ZEUS}} \; ,
    \end{array}
\end{displaymath}
are in good agreement with the value measured at TEVATRON:
$M_W = 80.33 \pm 0.15 \; GeV$ \cite{TEVATRON}, where the $W$ boson is produced
on mass shell. The data used in these determinations are from the 1993 run
with an electron beam energy of $ 26.7 \; GeV$ and the 1994 run with an
electron and positron beam energy of $ 27.5 \; GeV$. The collected integrated 
luminosities were: $0.33 \; pb^{-1}$ (H1) and $0.55 \; pb^{-1}$ (ZEUS)
for $e^-p$ scattering in 1993 and $ 0.36 \, (2.70) \; pb^{-1}$ (H1) and
$ 0.27 \, (2.93) \; pb^{-1}$ (ZEUS) for $e^-p$ $(e^+p)$ scattering in 1994.
The integrated luminosities used in these analysis are still
very small and the precision of the $W$-mass measurements should improve when
more data are analyzed. In any case, since the $W$ only enters in a 
$t$-channel exchange in deep inelastic processes and the cross-section
for on shell $W$ production is very small \cite{BAUR}, one cannot expect
to achieve at HERA the precision obtained at TEVATRON \cite{PROC}, but 
a consistency check is still interesting.

The effects of new physics are often parametrized via effective, dimension
$6$ four-fermion interaction terms (contact terms) which are added to the 
Standard Model (SM) Lagrangian. These terms can be originated either via an 
exchange of a heavy particle or as a result of a possible composite 
nature of the fermions involved. In the first case, the mass scale 
appearing in the contact term is interpreted as the mass of the exchanged 
particle over the coupling constant, while in the second case it is related 
to the compositeness scale in the strong coupling regime \footnote{See 
\cite{WUDKA} for a recent discussion on the physical interpretation on the
effective operators.}. 

The discussion on the effects of four-fermion contact interactions at HERA
began with the first studies of the HERA physics potential, long before the
collider was built \cite{VARIOS}. The excess of events above SM
predictions in $e^+p \to e^+$jet at high $Q^2$ recently observed by H1 
\cite{H1P} and ZEUS \cite{ZEUS2} has prompted a renewed interest on this subject
\cite{ALTAR,BABU,BARGER,BARTOLOMEO,AKAMA,BUCHMULLER}.
 All these studies, including the
old ones, concentrate on neutral current processes and only recently Altarelli
and collaborators \cite{ALTARELLI} have considered the effects of contact 
terms in charged current processes in relation with the neutral current 
excess of events. In this paper we
will obtain bounds for the $e \nu q \bar{q}^\prime$ contact terms 
contributing to charged current deep inelastic scattering from the 
published data for the $Q^2$ and $x$ distributions in
$e^\pm p \to \nu^{^{\hbox{\hspace*{-3mm}{\tiny(---)}}}} X$ 
\cite{H1,ZEUS,H1P}.  

Low energy effects of physics beyond the SM, characterized by a
mass scale $\Lambda$ much larger than the Fermi scale, can be studied by a
non-renormalizable effective lagrangian, in which all the operators are
organized according to their dimensionality. Since the energies and momenta
that can be reached in present experiments are much lower than $\Lambda$, it
is expected that the lowest dimension operators provide the dominant
corrections to the SM prediction. The relevant lagrangian for
$ep$ scattering, including dimension $6$, four-fermion operators is:
\begin{equation}
{\mc L} = {\mc L}_{SM} + {\mc L}_V + {\mc L}_S,
\end{equation}
where ${\mc L}_{SM}$ is the SM lagrangian and the two dimension
$6$ operators have been splitted into one term containing vector currents and
one containing scalar currents:
\begin{equation}
\label{LAG}
\begin{array}{lcl}
{\mc L}_V & = & \eta_{LL}^{(3),q} (\bar{l}\gamma_{\mu}\tau^{I}l)
               (\bar{q}\gamma^{\mu}\tau^{I}q) +
	       \eta_{LL}^{(1),q}(\bar{l}\gamma_{\mu}l)(\bar{q}\gamma^{\mu}q) \\
         & + & \eta_{LR}^{u}(\bar{l}\gamma_{\mu}l)(\bar{u}\gamma^{\mu}u) +
	       \eta_{LR}^{d}(\bar{l}\gamma_{\mu}l)(\bar{d}\gamma^{\mu}d) \\ 
         & + & \eta_{RL}^{q}(\bar{e}\gamma_{\mu}e)(\bar{q}\gamma^{\mu}q) \\
         & + & \eta_{RR}^{u}(\bar{e}\gamma_{\mu}e)(\bar{u}\gamma^{\mu}u) +
               \eta_{RR}^{d}(\bar{e}\gamma_{\mu}e)(\bar{d}\gamma^{\mu}d)  \\
{\mc L}_S & = & \eta_{d}(\bar{l}e)(\bar{d}q)+\eta_{u}(\bar{l}e)(\bar{q}u)+
               \tilde{\eta}_{u}(\bar{l}u)(\bar{q}e) + c.c. 
\end{array}
\end{equation}
\noindent
Where c.c.\ means conjugate terms, the $SU(2)$ doublets $l=(\nu,e)$ 
and $q=(u,d)$ denote left-handed
fields ($Ll = l, Lq = q, \hbox{with}\ L=\frac{1-\gamma_5}{2}$) and the SU(2)
singlets $e$, $u$ and $d$ represent right-handed electron, up-quark and
down-quark ($Re=e, Ru=u, Rd=d, \hbox{with}\ R = \frac{1+\gamma_5}{2}$).
It is customary to replace the coefficients $\eta$ by a mass scale $\Lambda$:
\begin{equation}
\eta = {\epsilon g^2 \over \Lambda^2},
\end{equation}
with $\epsilon = \pm 1$ taking into account the two possible interference 
patterns. $\Lambda$ is interpreted as the mass scale for new physics in the 
strong coupling regime, i.e.\ with
\begin{equation}
{g^2 \over 4 \pi} = 1.
\end{equation}
The experimental bounds for $\Lambda$ in ${\mc L}_V$ from direct searches 
in neutral current processes range
between $1.4 \; TeV$ and $6.2 \; TeV$ at $95 \% \; C.L.$, depending on the 
particles involved and the operator helicity structure \cite{PDG,OPAL}. More 
stringent bounds can be obtained from parity violating interactions in Cesium
\cite{LANGACKER}, however they can be avoided if the new physics possess some
global symmetries \cite{BUCHMULLER,NELSON}. None of these bounds apply 
to a $e \nu q \bar{q}^\prime$ contact terms, which are the ones that we will 
study in these letter. In the case of ${\mc L}_S$ there
are very strong bounds on $\Lambda_S$ from the ratio 
$R=\dis{\Gamma(\pi \to \mu \nu) \over \Gamma(\pi \to e \nu)}$ due to the fact 
that the operators in ${\mc L}_S$ do not lead to a helicity suppression in these
processes as the SM does \cite{ALTARELLI,SHANKER}. 
However, these bounds can be 
avoided if the new interaction is proportional to the fermion masses, as would 
be the case for the exchange of a heavy scalar, Higgs-like particle. 

The relevant terms for charged current processes are the first term from
${\mc L}_V$ and the whole ${\mc L}_S$. It is interesting to point out here that 
there are two terms from ${\mc L}_V$ contributing to neutral current processes
with an $LL$ helicity structure, while only one of them contribute
to charged current processes: the one proportional to $\eta_{LL}^{(3),q}$. 
Thus, both measurements are needed to completely study the 
effects of new physics.

The cross sections for $e^-p\to \nu X$ and $e^+p\to \bar{\nu} X$ can be obtained
in a straightforward way :
\begin{eqnarray}
\left( \frac{d^2 \sigma_{e^-p\to \nu X}}{dQ^2 dx} \right)_{CM} 
      & = & \frac{1}{\pi} \left( \frac{G_F}{\sqrt{2}} \frac{M_W^2}{Q^2+M_W^2} -
      \frac{\pi}{2} \frac{\epsilon}{\Lambda_{LL}^2} \right)^2
    \sum_{i=1}^{2} \left[ u_i(x,Q^2)+
    (1-y)^2 \bar{d}_i(x,Q^2) \right] \nonumber \\
& + & \frac{\pi}{16} \left( \frac{1}{\Lambda_d^4}+
     \frac{1}{\Lambda_u^4} \right) y^2 
     \sum_{i=1}^{2} \left[u_i(x,Q^2)+
     \bar{d}_i(x,Q^2) \right] \nonumber \\
& + &\frac{\pi}{16\tilde{\Lambda}_u^4}
     \sum_{i=1}^{2} \left[(1-y)^2u_i(x,Q^2)+\bar{d}_i(x,Q^2) \right] ,
     \label{electroncs}
\end{eqnarray}
\begin{eqnarray}
\left( \frac{d^2 \sigma_{e^+p\to \bar{\nu} X}}{dQ^2 dx} \right)_{CM} 
     & = &\frac{1}{\pi} \left( \frac{G_F}{\sqrt{2}} \frac{M_W^2}{Q^2+M_W^2} -
     \frac{\pi}{2} \frac{\epsilon}{\Lambda_{LL}^2}\right)^2
    \sum_{i=1}^{2} \left[ \bar{u}_i(x,Q^2) + 
     (1-y)^2 d_i(x,Q^2) \right] \nonumber \\
& + & \frac{\pi}{16} \left( \frac{1}{\Lambda_d^4}+
      \frac{1}{\Lambda_u^4} \right) y^2 
     \sum_{i=1}^{2} \left[\bar{u}_i(x,Q^2)+
     d_i(x,Q^2) \right] \nonumber \\
& + &\frac{\pi}{16\tilde{\Lambda}_u^4}
     \sum_{i=1}^{2} \left[(1-y)^2 \bar{u}_i(x,Q^2)+d_i(x,Q^2) \right] 
     \label{positroncs},
\end{eqnarray}
where we have neglected the interference between $O(p^6)$ terms. Since
the terms in  ${\mc L}_S$ are helicity changing there is no interference
between these terms and the SM. Thus, the bounds we will obtain
for $\Lambda_u$, $\Lambda_d$ and $\tilde{\Lambda}_u$ will be much smaller
than the ones for $\Lambda_{LL}$. The vector contact terms increase (decrease)
the cross-section when $\eta_{LL}$ is negative (positive), while the scalar
contact terms always increase the value of the cross-section with respect
to the SM prediction. It is obvious from Eqs.\ (\ref{electroncs})
and (\ref{positroncs}) that the effects will be larger for larger values of
$Q^2$. We illustrate these in Fig.\ 1 were we show the differential 
cross-section $\dis{{d \sigma \over d Q^2}}$ for 
$e^- p \to \nu X$ in the Standard 
Model and with the addition of vector contact term with 
$\Lambda_{LL} = 1 \; TeV$ and the two values of $\epsilon$. The 
differential cross-section $\dis{{d \sigma \over dx}}$ is also modified and the 
effects are larger for larger $x$ (see Fig.\ 2). Actually, the partonic 
cross-section violates unitarity with the introduction of the new terms 
and a form-factor has to be introduced to decrease the value of 
the cross-section at very high energy. However, since the quark 
distribution functions decrease very quickly for large $x$, we can 
neglect the effects of such form-factors. 

We have performed a $\chi^2$ fit to the $Q^2$ distributions published by the two
experiments at HERA: Table 1 in \cite{H1}, Tables 2 and 3 in \cite{ZEUS}
and Table 3 in \cite{H1P}. In the fits of the data from the first two references
we have included statistical and systematic errors, while for the third
one only statistical errors are included. The data from the third
reference has been converted into a distribution in bins of $Q^2$ in
such a way that it can be combined with the data from the other
references. We have always used the MRSA parton density parametrization
\cite{MRSA}, as the experimental groups have done in their fits to the 
$W$ mass. In our fit we have included the SM radiative 
corrections, but we have neglected the interference between 
these radiative corrections and the contact terms. Certainly, all the fits are 
compatible with the SM and only lower bounds on the mass scales
$\Lambda$ are obtained. These bounds do not change when we change the
$W$ mass within its experimental error. We have taken 
$M_W = 80.33 \pm 0.15 \; GeV$. The bounds we have obtained at $ 95 \% \; CL$
are shown in Table 1, where we have assumed a family independent contact term.
If only a contact term between members of the first family is allowed, the 
bounds are relaxed a $10 \%$ for the vector and a $2 \%$ for the scalar 
contact terms to those shown in Table 2. The bounds on $\Lambda_{LL}$ are a
factor $2.5$ lower than bounds obtained by the OPAL Collaboration \cite{OPAL}.
However, we should remind the reader at this point that the operators 
relevant in
neutral current processes are different than those in charged current
processes. Indeed, the latter involve only the term proportional to 
$\eta_{LL}^{(3)}$, i.e. the operator in which the currents transform as an
$SU(2)_L$ triplet, while neutral current processes, even for the $LL$
helicity combination, also receive a contribution from the operator involving
$SU(2)_L$ singlet currents. More definitely, in neutral current one is 
actually measuring the sum $\eta_{LL}^{(3)} + \eta_{LL}^{(1)}$. Thus, 
measurements on both processes are needed to completely fix the constants
in the lagrangian (\ref{LAG}).

The distribution $\dis{d \sigma \over dx}$ can also be used to obtain bounds
for $\Lambda$. In this case we have used the data from Table 4 in \cite{ZEUS}.
The results of the fit, see Table 3, are similar to those 
obtain from the $Q^2$ distributions. In particular, due to the presence of the 
interference between the vector contact term and the SM, the bounds on
$\Lambda_{LL}^-$ improve with respect to the ones obtained from the 
$Q^2$-distribution, while the bounds for $\Lambda_{LL}^+$ are somewhat weaker.
Since the effect of the scalar contact term is also an increase in the 
cross-section, as it is the case for the vector contact term with 
$\epsilon = -1$, the $x$-distribution is also more sensitive to the presence
of these terms.

The bounds obtained for the scalar operators from the HERA data are three orders of
magnitude lower than those obtained from pion decays \cite{ALTARELLI,SHANKER}.
In order to avoid these bounds we have also considered that the couplings 
of the dimension $6$ operators are proportional to the masses of the down-type
particles involved in the process. In particular, as an example, we have 
considered $\eta_S^s = 20 \eta_S^d$ and repeated the same fits. The results
from the $Q^2$ and $x$ distributions are shown in Table 4.

In summary, we have obtained bounds on the mass scales appearing in the 
dimension $6$, $e \nu q \bar{q}^\prime$ four-fermion contact 
interactions relevant for charged current
processes. This contact term has not been studied very much in the past getting
to the point that there is no entry for it in the Particle Data Group 
compilation on the search for compositeness \cite{PDG}.
Our bounds have been obtained from fits to the $Q^2$ and $x$
distributions measured at HERA by H1 and ZEUS. The $Q^2$-distribution provides
the strongest bound on a vector contact term with $\epsilon = 1$, while for
$\epsilon = -1$ and for an scalar contact term it is the $x$-distribution
the one that provides the strongest bounds. This is just an accident from
the set of data we have used. Both distributions are sensitive to the presence
of a contact term in a characteristic way which may be useful to pin down the 
origin of a possible departure of the data from the SM predictions. The lower 
bounds we obtain are lower than $ 3 \; TeV$, which is the value needed to 
explain
the excess of events observed in neutral currents. We agree, thus, with
the authors of Ref. \cite{ALTARELLI} on their conclusion that no effects
on charged current processes must have been observed if the anomaly in the
neutral current cross-section is due to a four-fermion contact term.

We thank W. Buchm\"uller for discussions at the early stage of this work
and F. del Aguila for a careful reading of the manuscript and his suggestions 
to improve it. This work is partially supported by CICYT under contract
AEN96-1672 and by Junta de Andaluc{\'\i}a. 

\newpage

\begin{center}
\begin{tabular}{||c||c|c|c|c||}   \hline
 
         &        &             &     ZEUS +      &           \\ 
         &  ZEUS  &  H1(93-94)  &    H1(93-94)    &  H1(94-96)\\ \hline\hline
$\Lambda_{LL}^{+}$   
         &  2.58  &    2.39     &      3.00       &     2.16    \\ \hline
$\Lambda_{LL}^{-}$ 
         &  3.38  &    2.39     &      3.38       &     5.01    \\ \hline
$\Lambda_{d}^{\pm}, \Lambda_{u}^{\pm} $ 
         &  0.63   &    0.67      &      0.70        &     0.81     \\ \hline
$\tilde{\Lambda}_{u}^{\pm}$
         &  0.77   &    0.72      &      0.79        &     1.07     \\ \hline
\end{tabular}

\begin{minipage}{11cm}
\small{Table 1: Lower bounds (in TeV) on contact term mass scales
obtained from various $Q^2$-distribution experimental data.
Universal contact terms are used.}
\vspace{0.25cm}
\end{minipage}

\begin{tabular}{||c||c|c|c|c||}   \hline
         &        &             &     ZEUS +      &           \\ 
         &  ZEUS  &  H1(93-94)  &    H1(93-94)    &  H1(94-96)\\ \hline\hline
$\Lambda_{LL}^{+}$ 
         &  2.34  &    2.20     &      2.72       &     1.95    \\ \hline
$\Lambda_{LL}^{-}$ 
         &  3.00  &    2.24     &      3.11       &     4.58    \\ \hline
$\Lambda_{d}^{\pm}, \Lambda_{u}^{\pm}$ 
         &  0.59   &    0.66      &      0.68        &     0.79     \\ \hline
$\tilde{\Lambda}_{u}^{\pm}$
         &  0.74   &    0.69      &      0.77        &     1.02     \\ \hline
\end{tabular} 

\begin{minipage}{11cm}
\small{Table 2: Lower bounds (in TeV) on contact term mass scales
obtained from various $Q^2$-distribution experimental data.
Only first family contact terms are used.}
\vspace{0.25cm}
\end{minipage}

\begin{tabular} {||c||c|c||}  \hline
                  & universal    &   first family   \\ \hline\hline
$\Lambda_{LL}^{+}$  
                  &   2.24      &    2.05       \\ \hline
$\Lambda_{LL}^{-}$ 
                  &   4.58      &    3.96        \\ \hline 
$\Lambda_{d}^{\pm}, \Lambda_{u}^{\pm}$ 
                  &   0.72       &     0.68          \\ \hline
$\tilde{\Lambda}_{u}^{\pm}$ 
                  &   0.86       &     0.81          \\ \hline
\end{tabular}

\begin{minipage}{11cm}
\small{Table 3: Lower bounds (in TeV) on contact term mass scales
obtained from the $x$-distribution data measured by ZEUS \cite{ZEUS}.}
\vspace{0.25cm}
\end{minipage}

\begin{tabular} {||c||c|c|c|c|c||} \hline 
 &        &             &     ZEUS +      &            &  ZEUS  \\ 
 &  ZEUS  &  H1(93-94)  &    H1(93-94)    &H1(94-96) &(x distribution)\\ \hline\hline
$\Lambda_{d}^{\pm}, \Lambda_{u}^{\pm}$ 
 &  1.92  &    1.59      &      1.92        &    1.98   &  2.24   \\ \hline
$\tilde{\Lambda}_{u}^{\pm}$ 
 &  2.34  &    1.87      &      2.34       &     2.64   &  2.64   \\ \hline
\end{tabular} 

\begin{minipage}{11cm}
\small{Table 4: Lower bounds (in TeV) on scalar contact term mass scales
obtained from the experimental data on $Q^2$-distribution (first four columns)
and $x$-distribution (last column).
We have assumed $\eta_S^s = 20 \eta_S^d$.}
\vspace{0.25cm}
\end{minipage}

\end{center}

\newpage

\begin{center}
{\bf Figure Captions}
\end{center}

\noindent
\begin{description}
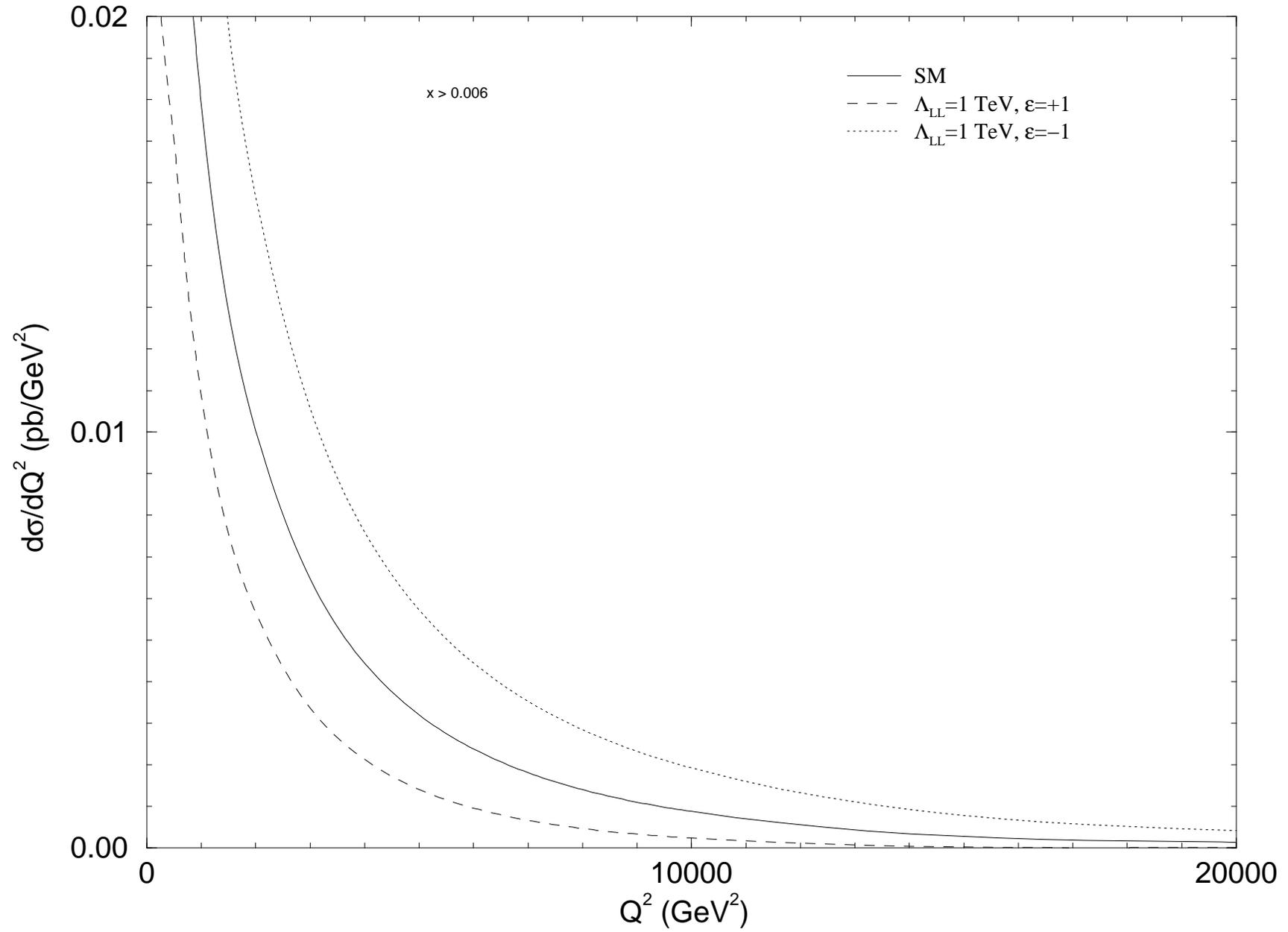
\item[Fig. 1:] $Q^2$-distribution for $e^- p \to \nu X$ in the Standard Model
(solid line) and with a vector four-fermion contact term with $\Lambda_{LL}
= 1 \; TeV$ and $\epsilon = 1$ (dashed line) and $\epsilon = -1$ (dotted
line). In all the cases a cut $x>0.006$ has been applied.

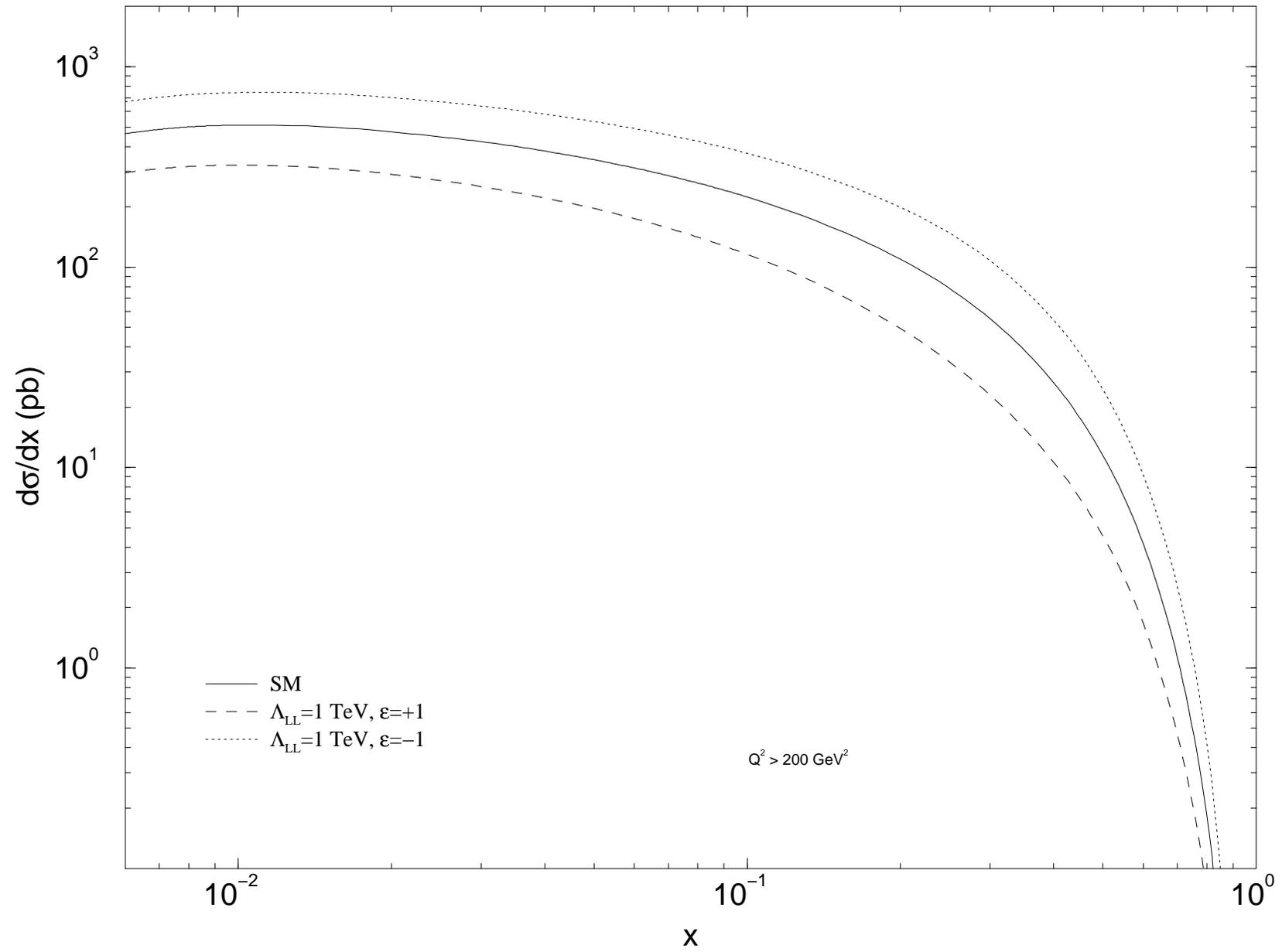
\item[Fig. 2:] $x$ distribution for $e^- p \to \nu X$ with the same conventions
as in the previous figure. Now a cut $Q^2 > 200 \; GeV^2$ has been applied.
\end{description}

\end{document}